\documentclass[referee]{aa} 
\usepackage{graphicx}
\usepackage{amsmath}
\usepackage[normalem]{ulem}

\begin{document}
   \title{Evidence of intense hot ($\simeq$340K) dust in 3CR radio galaxies}

\subtitle{The most dissipative source of  cooling in AGNs}
 
\authorrunning{B. Rocca-Volmerange \& M. Remazeilles}
\titlerunning{Hot dust in 3CR radio sources}
   \author{B. Rocca-Volmerange
          \inst{1,2}
          \and
          M. Remazeilles \inst{1} }
   \offprints{brigitte.rocca@iap.fr}
\institute{Institut d'Astrophysique de Paris,
	98bis Bd Arago, F-75014 Paris
      \and Universit\'e de Paris-Sud XI, I. A. S., 91405 Orsay cedex, France}

\date{Received ; accepted }  
\abstract{The spectra of the powerful 3CR
radio  galaxies present  a  typical distribution  in the  far-infrared
(FIR).  From the observed  radio to X-ray spectral energy  distribution (SED)
  templates, we  propose to
subtract  the   typical  energy  distributions   of,  respectively,  the
elliptical galaxy  host and  the synchrotron radiation.  The resulting
SED reveals that  the  main dust emission 
 is well
fitted  by the  sum  of  two blackbody  components  at the  respective
temperatures 340K $\pm$  50K and 40K$\pm$16K.  
 When the AGN is active,
the  energy rate released  by hot dust  is much  more dissipative
than cold dust and stellar emission, even when the elliptical galaxy 
emission is maximum at age of $\simeq$ 90 Myr. Hot dust appears
as a huge  cooling source which implies an  extremely short time-scale
$t_{cool}$.   In  balance,  with  the  short  gravitational  time-scale
$t_{grav}$~  of massive  galaxies, the  dissipative self-gravitational
models (Rees \&  Ostriker, 1977) are favoured for  radio sources. They
justify   the   existence  of   massive   radio   galaxies  at   $z$=4
(Rocca-Volmerange et al., 2004).  The synchrotron emission is emitting
up to the X-ray wavelength  range, so that strong "EXOs" sources could
be assimilated to 3CR radio sources.  This analysis applied to ISO and
SPITZER  data on  a  larger sample  will  statistically confirm  these
results.
} 
\maketitle

\section{Introduction}
In  Rocca-Volmerange et  al.,  2004, hereafter  RV04,  we showed  that
powerful  radio  sources  are  hosted by the  most massive  galaxies.  Based  on
measurements  of  stellar masses  with  robust  evolution models,  the
maximum   mass   is   limited    by   the   fragmentation   limit   at
$\simeq$10$^{12}$ M$_{\odot}$~  (Rees \& Ostriker,  1977, Silk, 1977)
clarifying the interpretation of the so-called $K$-$z$~relation in the 
$K$-band Hubble diagram.
Moreover in this diagram, the authors give a constraint on galaxy types:
only host galaxies of elliptical type
fit the radio galaxy distribution from $z$=0 to 4.  However at $z$=4, the time-scale
of  mass accumulation  becomes so short that to form massive galaxies requires  short  dissipative time
scales, of the  same order as gravitational time  scales. 

Typical UV
to  radio SEDs  of 3CR  galaxies were
compiled  from observations of ISO,  IRAS  and IRAM  
observatories  by Andr\'eani et  al.,
(2002).   From  the FIR emission,  the  authors conclude  that there is 
 double
emission  from a  dusty  torus   and  a  larger-scale   (cooler)  dust
distribution in the host galaxy. We propose more details on the 
dust temperatures by a multi-component approach including stars
and jet. 

Statistically well identified by  their high radio power, 3CR galaxies
 are also  host by massive stellar populations,  contributing to the
 optical and  infrared emissions.   In the near-infrared,  the stellar
 emission  of radio  sources is  often similar to  populations of
 massive  elliptical  galaxies.    In  the  mid-infrared,  the  recent
 analysis of a significant sample of early type galaxies observed with
 ISOCAM (Xilouris et al, 2004) shows that the emission is dominated by
 the presence of  the PAH feature at 6.7$\mu$m, an  excess of hot dust
 at 15$\mu$m and  a cold component at ~30-40K.   The comparison of the
 FIR emission of elliptical galaxies with that of radio sources will give 
 specific information on the  AGN contribution. Using templates from the
 code P\'EGASE (www.iap.fr/pegase) we are able to predict the 
 stellar emission  at all  galaxy ages,  while  the synchrotron
 radiation contributes to SEDs  when the AGN  is
 active.

Another objective is to estimate the cooling time-scale at  
the earliest phases of galaxy formation. 
Instead of using the classical cooling function of only helium  and hydrogen
clouds, each dissipative source (stars, gas, dust and AGN) has to be 
individually considered during galaxy evolution.   In section 2, we
compare the various components 
to the averaged observed SED of 3CR radio galaxies. Section 3 presents the
best fit of dust emission with the sum of two main blackbody laws, the hot one
is an intense source of  dissipation. Section 4
predicts the  dissipation rate from various  sources (synchrotron, stars,
gas and dust),  to be compared to the dynamical  time scale.  The last
section gathers the discussion and conclusion.
 
\section{Stellar and synchrotron emissions} 
 The striking similitude of the radio to  UV SEDs of 3CR radio galaxies and
quasars (see Fig.1 to  4 in Andr\'eani et al,  2002) evokes similar
properties of the various components of these complex systems.
All the spectra present
a typical gap from the  radio emission for $\lambda < $1mm, presuming
similar  properties of  dust in  3CR  galaxies.

\subsection{Stellar  and nebular components}
\begin{figure*}
\centering
\includegraphics[width=13cm]{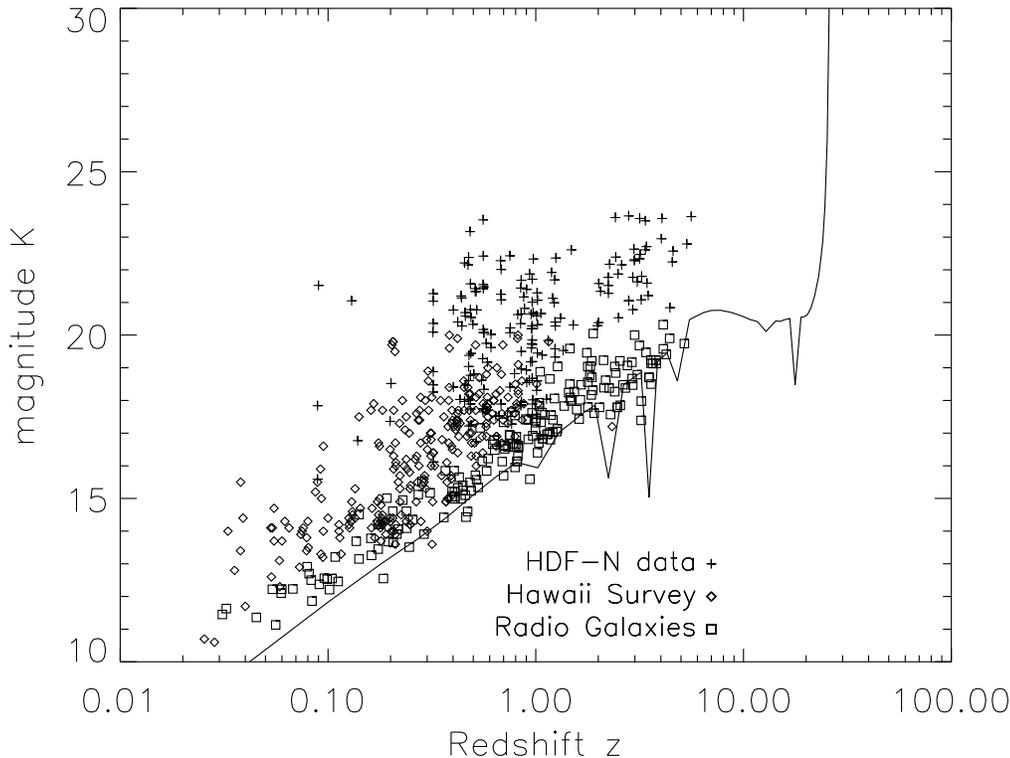}
\caption{The distribution of radio (squares) and field (diamonds, crosses)
 galaxies 
in the $K$-band Hubble diagram, compared to the predicted sequence of elliptical
galaxies of masses 10$^{12}$M$_{\odot}$, the adopted 
redshift of formation is $z_{for}$=30. Sequences with  
other masses and redshifts of formation
were already tested in Rocca-Volmerange et al. 2004. The main emission lines from AGN appearing in the
$K$~band for 0$< z <$ 30 (H$_{\alpha}$, [OIII]5007$\AA$~ + H$_{\beta}$, 
[OII]3727$\AA$~and Ly$_{\alpha}$1215$\AA$) are also plotted with 
width of 10$\AA$. 
Previous conclusions are unchanged: 10$^{12}$M$_{\odot}$~ 
is the upper mass limit of galaxies and also the fragmentation limit
predicted by models}
\label{figure:Kz30}
\end{figure*}

Radio galaxies  are embedded in  massive elliptical galaxies,  even at
high redshifts (van Breugel et  al, 1998, Lacy et al, 2000, Pentericci
et al, 2001): from high  luminosities L$\simeq $ 3 to 7 L$_*$(Papovich
et al, 2001  ), they may reach up  to 10 L$_*$ (Mc Lure  et al, 2004).
The 3CR radio galaxies are  the most powerful galaxies in the $K$-band
Hubble   diagram,   limiting  the   distribution   by  the   so-called
$K$-$z$~sequence. In RV04, we  checked with the evolution model P\'EGASE
that  only  elliptical  scenarios  of  10$^{12}$M$_{\odot}$~  baryonic
masses are  able to  explain the distribution  of the  brightest radio
galaxies   in    the   Hubble   diagram. 
 However huge emission lines are typical of massive radio galaxies
while elliptical galaxies show no emission lines. To confirm our mass 
estimates, we also include in our modeling 
 the nebular emission of gas ionized by the AGN component
(Moy \& Rocca-Volmerange, 2002), computed with the code CLOUDY (Ferland, 1996).
 In    the   present   paper
(Fig. \ref{figure:Kz30}), the observed radio galaxies in the 
Hubble $K$~ diagram are well fitted by 10$^{12}$M$_{\odot}$ elliptical 
models and AGN emission lines. The emission line widths are assumed to be  
10$\AA$~ at $z$=0. The redshift of
elliptical galaxy formation is  $z_{for}$=30,  instead  of  
 $z_{for}$=10  in   RV04.  Our
conclusions  on galaxy  formation limited  by the  fragmentation limit
10$^{12}$M$_{\odot}$~remain  unchanged,  allowing us to adopt this 
mass for 3CR powerful radio galaxy hosts. The NIR predictions 
from 1 to 5 $\AA$~ are strongly dependent on the modeling 
of cold star populations. The effective temperatures of giant branches
and asymptotic giant branches, as well as the mass population density
are crucial but not 
accurately estimated.  We check the P\'EGASE elliptical
modeling of the two populations by comparing predictions to the 
observational template (see Fig.2 in Fioc \& Rocca-Volmerange, 1996).
However, the separation of the stellar and hot dust component
will be less robust between $\lambda$=3 to  5$\mu$m than for  
$\lambda >$ 5$\mu$m.

Most scenarios of galaxy evolution take into account 
the extinction process by dust, computed with a transfer model 
in ellipsoidal or slab geometries (Fioc \& Rocca-Volmerange, 1997). 
However in the elliptical scenario, galactic winds expel gas and dust 
at 1 Gyr, so that at $z$=0 our modelling only simulates the stellar emission.
In  the following,  we  shall
adopt  the  SED model  of  elliptical  galaxies  for predictions of the
underlying populations of radio sources. 

\subsection{Synchrotron radiation}
The synchrotron radiation in radio galaxies follows a power law 
F$_{\nu}$ = $\nu^{\beta}$. Observations in the radio domain 
are well fitted by $\beta$ = -1.04 (Andr\'eani et al., 2002).

\begin{figure*}
\centering
\includegraphics[width=13cm]{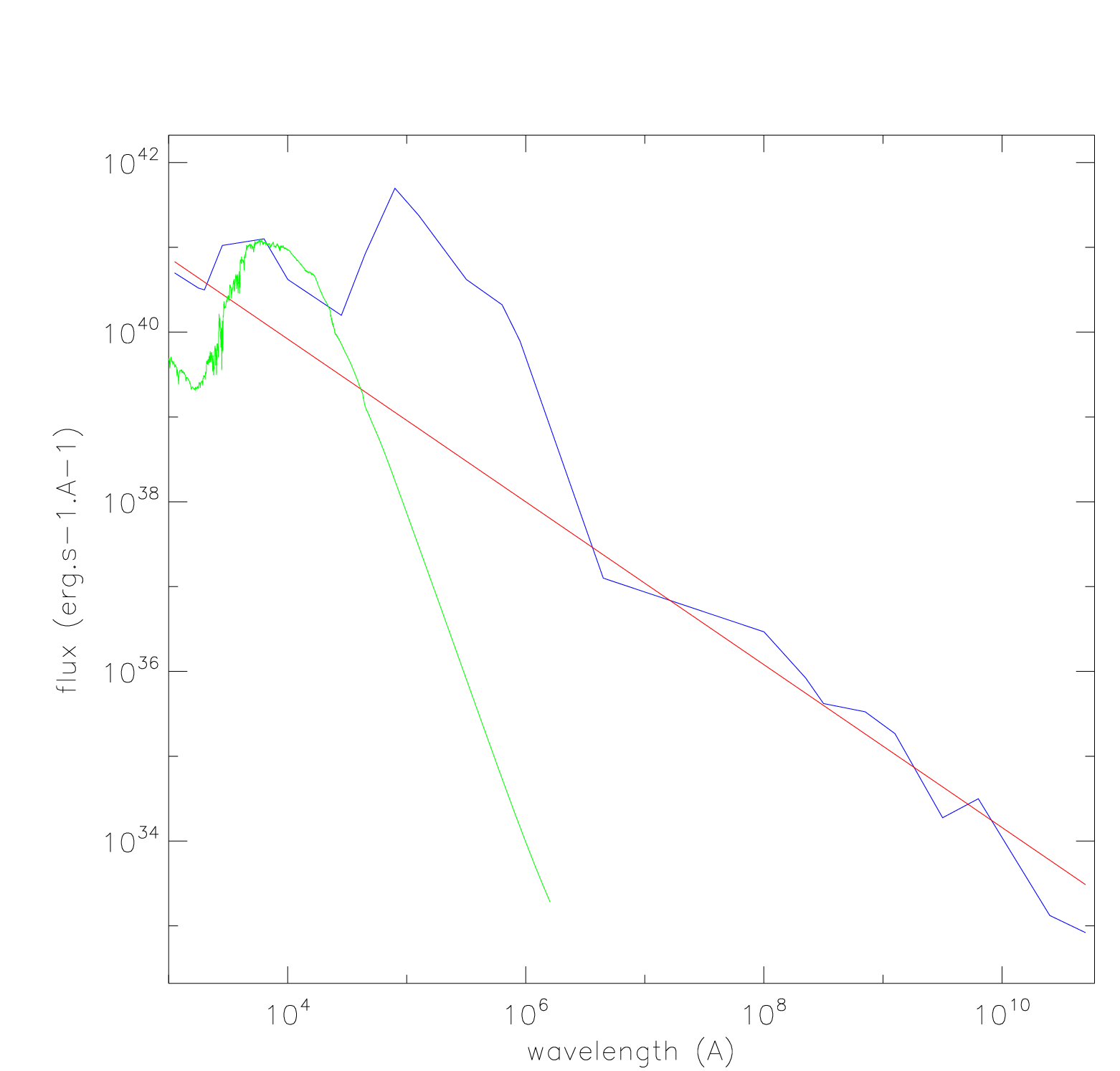}
\caption{The observed energy distribution of 3CR galaxies by Andr\'eani et al, 2002 (full blue line) 
on the large wavelength range ($\lambda$ = 10$^{3}$~to 10$^{10}$ \AA~ is 
plotted
with the stellar SED of elliptical template (dotted green line) from P\'EGASE
and the synchrotron power law (full red line). The synchrotron radiation might contribute to 
the X-ray emission (see discussions on the so-called $EXOs$~ sources below). }
\label{figure:composantes}
\end{figure*}

The energy distribution and intensity of the dust emission appear
from the radio-to-UV SED template of 3CR radio galaxies:
 by comparing
the stellar  and synchrotron  emissions to the 
composite spectrum of averaged 3CR galaxy SEDs, we discover a large unresolved bump  from $\lambda$= ~ 1 to 500 $\mu$m (log($\nu_{Hz}$) $\simeq$12  to 15).  The
average redshift $z$=0.5  of the observational sample has been
corrected to the rest frame. The comparison is presented on Fig.\ref{figure:composantes}.

\section{Main hot and cold dust components}
\begin{figure*}
\centering
\includegraphics[width=13cm]{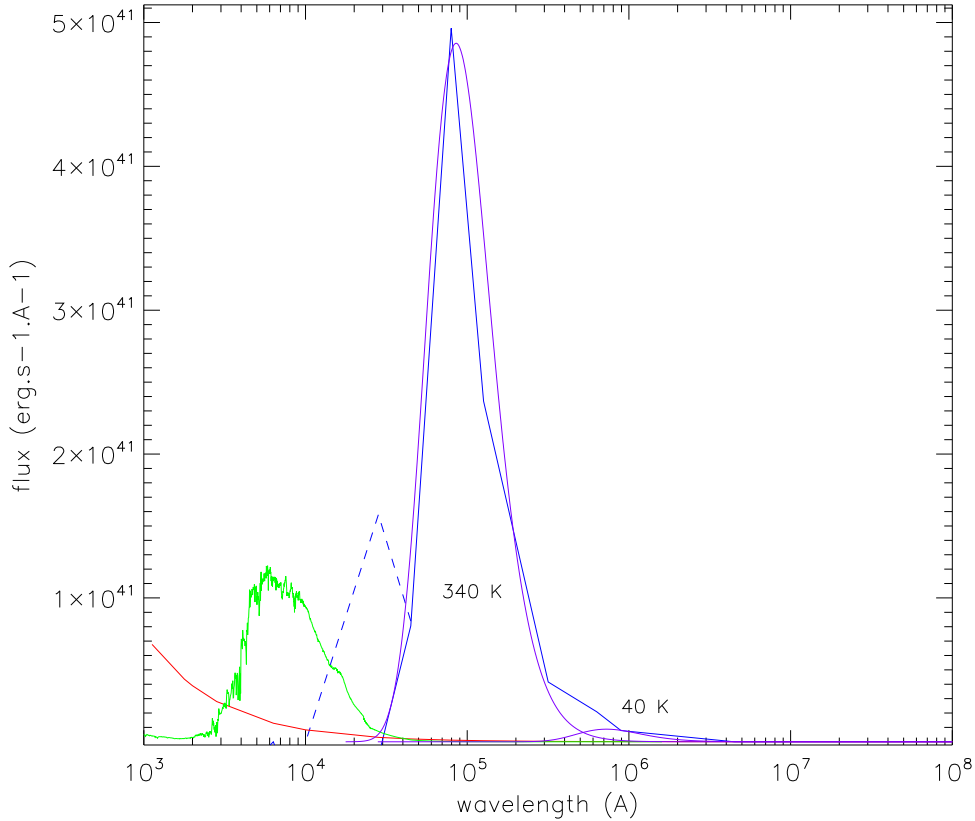}
\caption{The main three fluxes contributing to the radiative energy 
of the 3CR radio galaxy sample are the elliptical SED (green line), the synchrotron 
power law (red line) and the two blackbody laws (pink lines ) of respectively 
340K$\pm$50K and 40K$\pm$16K. The dust emission (blue lines) is derived from 
observations after subtraction
of stellar + synchrotron emissions.  For $\lambda >$ 3$\mu$m, the fit with the sum of 
two blackbody laws is robust (full line), while for $\lambda <$ 3$\mu$m,
the peak at the temperature $\simeq$~900K is more unaccurate (dashed line).}
\label{figure:temperatures}
\end{figure*}

The various grains  found in  the interstellar medium (Desert  et al.,
1990) contribute  at different levels  to the total FIR  emission of
radio sources. The global FIR emission is described as the superposition
of blackbody (BB) laws depending on the grain temperature, size and  nature
distribution.
We do not consider large lines  of PAH, generally attributed as signatures 
of star formation: in evolved elliptical galaxies, the star formation
activity is null or very faint. The first result of the analysis is that 
one unique BB 
temperature is unable  to reproduce the IR bump of dust deduced from
stellar and synchrotron corrections.  We then propose to limit the 
decomposition to only two BB laws. The second result is the excellent
fit for two extreme  temperatures: the hot component  
corresponds to the BB temperature of 340$K \pm$ 50$K$, 
the cold component to the BB temperature of 40$K \pm$ 16$K$.
Fig. \ref{figure:temperatures} presents
the stellar and
synchrotron fluxes F($\lambda$)as function of the wavelength $\lambda$~
and the two BB laws compared to the observed bump.
The hot wavelength peak derived from the Wien law at 340K is 8.5 $\mu$m. 
Fluxes at mid-height correspond to an accuracy of $\pm$50K.   The hot
component is an extreme source of energy dissipation. A similar temperature 
was observed in the far-infrared continua of quasars (Wilkes et al. 1998)
but quasars do not allow to disentangle the contribution of 
other components, in particular stellar. Also found in radio galaxies, 
this hot component may be attributed to the active galaxy nucleus (AGN): however the
hot dust component at 260 K found in early-type normal galaxies
from ISOCAM data (Ferrari et al., 2002) could be  of similar nature, even if 
it is strikingly less luminous. 

The cold wavelength peak derived from the Wien law at 40K is 72$\mu$m. 
Fluxes at mid-height correspond to an accuracy of $\pm$16K.This  cold
component may be compared to the already known 20K-60K
temperatures found 
by  {\it ISO} and {\it SPITZER} telescopes in  dusty evolved galaxies (Blain  et  al,
2004, Xilouris et al, 2004). 
Qualitatively  this result of two components  is confirmed by the  splendid
IRAC/SPITZER image of the  nearby radio galaxy Centaurus A 
(Fazio et al, 2004). 
 A third emission peak at $\simeq$~3 $\mu$m is possible from data 
(dashed line on Fig. \ref{figure:temperatures}).
The disentangling of the cold stellar and highly hot ($\simeq$900K) dust
emission  is less robust and of minor energy importance.

\section{The balance of energy  dissipation rates}
The classical cooling function predicted for initial clouds of hydrogen and helium
(Rees \& Ostriker, 1977 and references therein) becomes insufficient,
and may be wrong, when intense cooling sources are activated during galaxy formation. While stellar and cold dust emissions, in particular in case of rapid metal enrichment,
depend on star formation rate, the
processes related to the AGN are only dissipative 
when the AGN is active. However the cooling processes efficiently 
contributes to the decrease 
the cooling time-scale t$_{cool}$. We propose to estimate
hereafter the radiative energy balance, during the star formation
evolution when the AGN is active. We integrated all the previously considered
fluxes (stellar SEDs, BB laws, power law)
 on their largest  wavelength domain of emission. The dissipative energy rate 
$dE/dt$~ in ergs. s$^{-1}$~ is then computed for all sources. 
For simplicity, we separate stellar and AGN sources. Stellar sources
 evolve on the galaxy time-scale (14 Gyr) with passive and active 
evolution while AGN, supposed to be active for an arbitrary time
scale, is of short 
lifetime duration ($<$10$^{8}$~ years).

\begin{figure*}
\centering
\includegraphics[width=13cm]{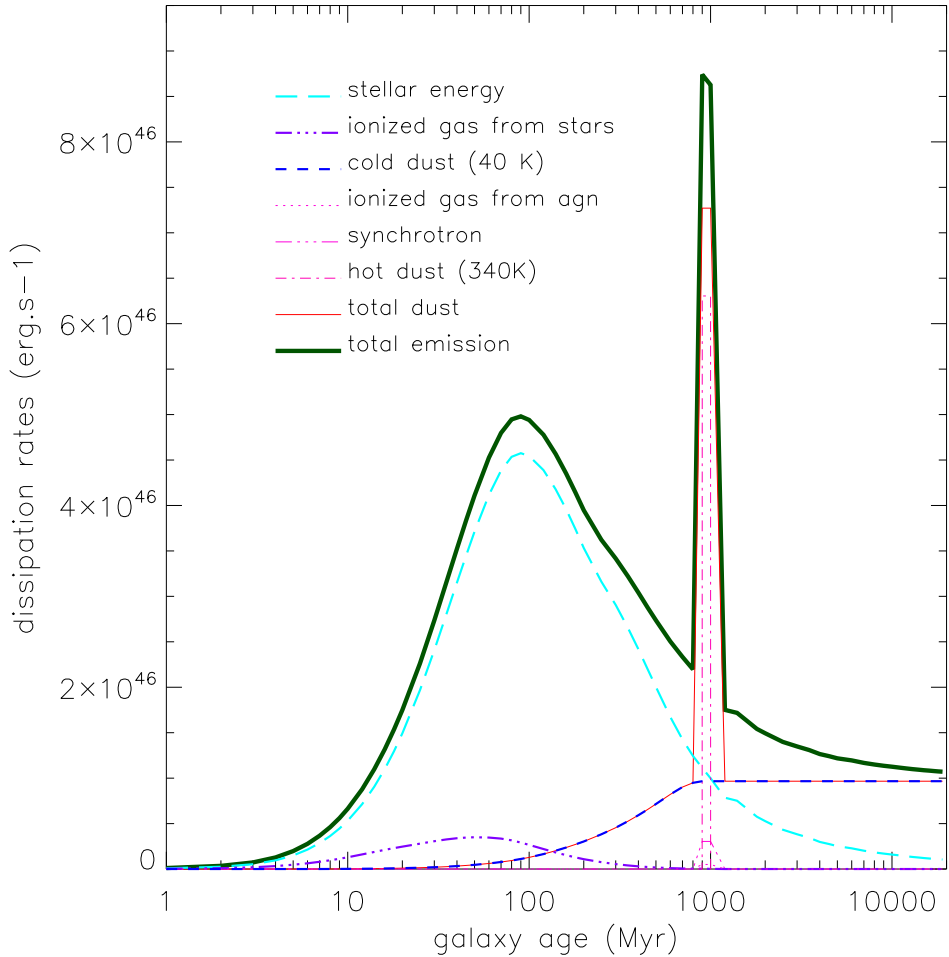}
\caption{Various stellar (blue color) dissipative rates $dE/dt$~ in  
erg.s$^{-1}$~as a function of time: stellar continuum (dashed light blue
 line), ionized gas from massive stars(dash-dot blue line), 
cold dust fitted on metals (dashed blue line). Dissipative rates
due to the AGN formed at 1 Gyr (pink/red color): ionized gas from AGN 
(dotted clear pink line), synchrotron (dashed-double dot  pink line ), hot dust
 (dashed-dot red line). The total dust emission is the pink full line
and the sum of all rates is the full green line. 
Radiative energy by supernova explosions (except neutrinos), too faint,
is not plotted.}
\label{figure:bilan}
\end{figure*}

\begin {itemize}
\item The supernova rate is predicted at all ages from the 
evolution scenario of elliptical galaxies. Adopting the average 
luminosity of 10$^{42.5}$erg s$^{-1}$~from Nomoto models,
 the radiative energy rate
is a minor component of the global emission. In fact, the largest fraction of
the explosion energy heats the interstellar medium, up to the escape 
velocity.
For more detailed predictions, the neutrinos could be taken into account, 
so that our results are a lower  limit.
\item From the gas ionized by massive stars, hydrogen and oxygen 
lines are the strongest sources of dissipation through the  main lines 
(Ly1215\AA, H$\alpha$, H$\beta$, 
[OIII]5007\AA, [OII]3727\AA). The evolution
of the number of Lyman continuum photons N$_{Lyc}$~is 
predicted by P\'EGASE and the emission lines are those of 
a classical HII region at the electronic 
temperature 8000K. The line intensities are then derived by 
taking into account the metallicity evolution of the gas. 
\item Huge envelops of  gas ionised by the AGN are observed at all $z$~
in the most powerful radio galaxies, in particular around the most distant ones
(van Ojik et al., 1997). Emission lines due to AGNs are computed with the 
code CLOUDY, assuming solar metallicity, the photo-ionisation parameter
log U = -2 and the density 100 cm$^{-3}$. 
\item The dissipative energy of stellar emission is computed at all ages 
with the bolometric luminosity as a function of time, a  standard output of the code P\'EGASE.

\item The  synchrotron radiation, mainly efficient in  active AGNs,
contributes  to a fraction of the  total radiative
energy. The power law is integrated  on  the whole
energy range from X-rays to radio.

\item  The hot  and cold  dust  components, as found 
in the previous section, dissipate energies at rates $dE/dt$,
deduced from integration  of the  blackbody laws  on  their respective
wavelength  domains.  
\end{itemize}
 We  plot all  the  dissipation rates  on  Fig. \ref{figure:bilan}  by
adopting an active  AGN at 1 Gyr.  The hot  dust component is luminous
when the AGN is active while  the cold component, also found in elliptical
galaxies,  is supposed  to follow  the  evolution of  the host  galaxy
metallicity.  As a result, when the  AGN is  active, the  most important  source of
dissipation is  the hot dust component  at $\simeq$340K.

\section{Discussion and conclusion}
Based on SEDs of 3CR radio galaxies observed from the X-ray to radio
domain,
we identify the dust emission by subtraction of the template of
elliptical galaxies and of the synchrotron power law. 
Two main dust BB emissions  (and a possible minor hotter one) are revealed in the  
$F(\lambda)$-$\lambda$~diagram, while the classical diagrams 
$F(\nu)$-$\lambda$~ (or $\nu$ F($\nu$)- $\lambda$) (Haas et al, 1998) are less adapted 
to the component separation. 

The hot dust emission peaks at 8.5$\pm$3$\mu$m with a blackbody law of
340K$\pm$50K.   This component is in agreement  with the
standard AGN molecular torus model (Pier \& Krolik, 1993), embedded in
a cooler component. By comparing to the IRAC/Spitzer image 
of Centaurus A (Fazio et al, 2004) the bright hot  grain structure 
is not in the inner core but within a larger structure. 
Many speculations on the origin of the hot grain origin are possible,
we only conclude on the presence of a large
amount of dust photo-heated by the AGN. 

 The hot BB temperature uncertainty at the maximum peak 
includes errors due to the 
calibration and to the modeling of the stellar emission. From 
Fig \ref{figure:temperatures}, the stellar emission is negligible 
at about 8 $\mu$m. The large bar of $\pm$50K means that the resolved 
signatures of Polycyclic Aromatic hydrocarbons (PAH) at respectively 
7.7$\mu$m and 8.6$\mu$m could be included in this peak.

The most important uncertainty concerns the 3 $\mu$m zone, highly sensitive
to the subtraction. Moreover from observational data (Andr\'eani et al, 2002),
this zone is highly dispersed. The authors present the extreme
value for quasars which would indicate a strong emission peak. 
For radio galaxies, the $\simeq$~3 $\mu$m peak is significantly 
less energetic.

The main difference
with  previous  studies  on active galaxies  concerns  the  star
formation.  While the  starburst activity is preferentially researched
in the Ultra Luminous Infra  Red Galaxies (ULIRG) (Genzel \& Cesarsky,
2000), star formation is a minor component in  radio
sources, dominated by the evolved population of elliptical galaxies. 
However the 8.5$\pm$3$\mu$m  peak is also found in the
ULIRGs (Ultra  Luminous Infra Red  Galaxies). In general attributed  
to the PAH, the peak 
is not thermal and due to the episodic excitation by PAHs.  
It is still premature  to
conclude on the  link between the 8.5$\mu$m peak  discovered in strong
ULIRGs  and the  8.5$\mu$m peak  in powerful  radio galaxies. 
But the similitude of the two peaks deserves further analyses 
on hidden AGN in starbursts
or/and on star formation in AGN environments.

The  cool component at 40K is at about the same temperature than in
elliptical  galaxies (Xilouris et al, 2004). So that the origin is not 
necessarily linked to the presence of the AGN but to 
dusty populations of stars (low mass AGB stars or others).
Moreover because stellar winds in elliptical galaxies must eject gas and dust,
we need to justify the presence of dust.
Would it be possible that stellar winds expel 
the interstellar gas, while the denser and 
more embedded dust component is maintained in the galaxy centre environment? 
Grains, more massive than gas, could be 
more rapidly attracted towards the galaxy centre than the gaseous component,
the process may also depend on the angular momentum value. 
The time scale of grain infall towards the center is then shorter than the time scale 
of interstellar gas heating. 

Are these results on 3CR radio galaxies acceptable for all AGNs? 
3CR radio galaxies are the most powerful and massive galaxies 
hosting super massive black holes (McLure  \&
Dunlop, 2002). If dust emission, in particular hot dust, is due to
a collapsing process of dust towards the centre, the dust origin 
is not due to AGN which only heat grains and make them luminous. 
It could also be linked to a star formation
process if massive stars are totally embedded in dusty clouds since
no evidence of SED signatures due to starbursts is revealed in
the optical part of the SED. 
Whatever the origin of the energetic photons 
heating the hot dust structure, the 
energy released by this structure is 
considerable. An approached dynamical scale
$t_{grav}$ = 1/{(G $\rho$)$^{1/2}$ } $\simeq$ 600 Myr becomes 
comparable to the cooling time scale during the AGN phase. 
The total energy dissipated by the 340K emission, integrated
on the AGN duration (10$^{8}$ yr) gives a time scale
$t_{cool} >$ 400 Myr. 
Our evaluations
show that $t_{grav}$~ and $t_{cool}$~are 
quite comparable and the dissipation may regulate the self-gravitational
collapse models of galaxy formation
(Rees \& Ostriker, 1977).
The dissipation factor may be lower at higher redshifts 
when metals and dust mass are significantly 
lower. However, in case of a massive initial gas reservoir, 
the collapse is extremely rapid, the metal enrichment follows
and finally the grain emission is dominant.  
Another source of uncertainties is the energy released by neutrinos
from supernova explosions, so that the presence of the active nucleus
could not be necessary to dissipate energy at a low time scale comparable
to gravitational time scale. More detailed 
observations from the ISO archives and the rapidly increasing data set from the 
satellite SPITZER are required.

\begin{acknowledgements}
We thank Nick Seymour for reading the manuscript 
\end{acknowledgements}

\end{document}